\documentclass[12pt,preprint]{aastex}

\def\m{$\mu$m}
\def\ab{$\sim$}

\def\et{et al. }
\def\deg{\hbox{$^\circ$}}

\shorttitle{}
\shortauthors{Colbert \et}

\begin{document}


\title{The Bright Ages Survey. I. Imaging Data}


\author{James W. Colbert\altaffilmark{1}, Matthew A. Malkan\altaffilmark{2}, R. Michael Rich\altaffilmark{2}, Jay A. Frogel\altaffilmark{3}, Samir Salim\altaffilmark{2}, Harry Teplitz\altaffilmark{1}}
\altaffiltext{1}{Spitzer Science Center, California Institute of Technology,
    Pasadena, CA 91125}

\altaffiltext{2}{University of California, Los Angeles, CA, 90095}

\altaffiltext{3}{Ohio State University, Columbus, OH 43210}



\begin{abstract}

This is the first paper in a series  presenting and analyzing data for a K-selected sample 
of galaxies collected in order to identify and study galaxies at moderate to 
high redshift in rest-wavelength optical light. The sample contains 842 objects over 
6 separate fields covering 75.6 arcmin$^2$ down to K=20-20.5. We combine the K-band with 
UBVRIzJH multi-band imaging, reaching depths of R\ab26. Two of the fields studied also have
deep HST WFPC2 imaging, totaling more than 60 hours in the F300W, F450W, F606W, and F814W
filters. Using artificial galaxy modeling and extraction we measure 85\% completeness
limits down to K=19.5-20, depending on the field examined. The derived K-band number counts
are in good agreement with previous studies. We find a density for Extremely Red Objects 
(EROs; R-K$>$5) of 1.55$\pm$0.16 arcmin$^{-2}$ for K$<$19.7, dominated by the 1714+5015 field 
(centered on 53w002), with an ERO number density more than 3 times that of the other sample fields.
If we exclude the counts for 1714+5015, our density is 0.95+/- 0.14 arcmin. Both ERO densities are consistent with 
previous measurements due to the significant known cosmic variance of these red sources.
Keck spectroscopic redshifts were obtained for 18 of the EROs, all but one of which are emission galaxies.
None of the EROs in the 1714+5015 field for which we obtained spectroscopic redshifts are 
associated with the known $z$=2.39 over-density, although there are three different galaxy redshift pairs 
($z$=0.90, $z$=1.03, $z$=1.22).

\end{abstract}


\section{Introduction}

Great advances in mapping the high redshift universe have been made, largely through optically 
selected surveys, such as the Canada France High Redshift Survey\citep{lil96}, Hubble Medium Deep 
Survey \citep{ost98,rat99}, and the Deep Extragalactic 
Evolutionary Probe \citep{coi04,mad03} out to $z\approx 1$, and surveys of Lyman break-selected 
galaxies beyond z=2 \citep{mad96,ste99,ste04,ade05}. 
Luminosity density and star formation rate increase with redshift, reaching a maximum 
approximately at redshift 1 -- 2.5, what could be labeled as ``the Bright Ages''. 
Star formation density then remains relatively flat out through z=4, with evidence building for a drop out to z=6 
\citep{gia04,bun04,boa04} and beyond \citep{bob04}. 

One problem with optically-selected high-redshift surveys is that they are biased against the 
discovery of red galaxies and fall especially short in the search for old and/or reddened
star forming galaxies at $z>1.5$.  
A red galaxy must be an order of magnitude brighter in the 
rest wavelength optical than a blue galaxy to be included in typical optical high redshift 
surveys \citep[e.g.,][]{fra97}.  A new generation of deep near-infrared selected surveys were
undertaken with the aim of addressing this shortcoming.
The Las Campanas Infrared Survey \citep{mcc01}, K20 Survey \citep{cia02}, FIRES \citep{fra00},
and GOODS \citep{gia04}  are sensitive to rest-wavelength optical light even at high
redshift, mitigating the problems of reddening and sensitivity to passively evolving
populations at high redshift.

Surveys that base their selection on near infrared data
are capable of finding substantial numbers of red galaxies
at high-redshift
\citep{lab03,abr04,fon04}, including both dusty starbursts and evolved stellar populations.
There is mounting evidence that dust plays a significant role at high redshifts 
\citep{row97,sma02,elb02}, causing high extinction in the
rest ultraviolet; as a consequence many of these galaxies drop from optical surveys.
Ultraviolet flux is not a reliable star formation indicator, requiring
highly uncertain corrections \citep[i.e.,][]{cal97,dic97,tre98} and 
saturating at the highest luminosities \citep[\ab 10$^{10}$ L$_{\sun }$;][]{mar05}. 
Rest-frame ultraviolet is also typically dominated by recent starbursts, which can
be only a small percentage of the total galactic mass \citep[i.e.,.][]{sha05}. 

The commissioning of new telescopes and spectrographs has made the study of infrared-selected samples
a very active field. Our survey was constructed so as to take
advantage of archival HST imagery reaching to nearly the depth of
the Hubble Deep Field, as well as fields containing quasars and clustered CIV absorbers
with redshifts placing them in the target $z\sim 2$ bright ages interval.
Progress has been dramatic in this subject area, and in recent years we have seen
results from the K20 project \citep{mig05} with its survey area of
$\sim 52$ arcmin$^2$) and the GOODS survey \citep{gia04} of $\sim 300$ arcmin$^2$.

Our survey has distinct characteristics:
we have multicolor data in six widely separated fields, five of which have
evidence of clustering (either CIV absorption system or emission line galaxies)
at $z\sim 2.5$.   
Thanks to our photometric and spectroscopic redshifts, these
dispersed fields are studied over a wide slice of redshift space and their physical
separation makes our overall results less sensitive to cosmic variance while
giving an idea of how important clustering is to contributing to cosmic variance.
Further, our field size is well matched to followup with the
Keck LRIS spectrograph \citep{oke95} which has recently been upgraded
to achieve high sensitivity down to the atmospheric cutoff, a feature of special advantage
in measuring Ly alpha emission line redshifts in the $z>2$ Bright Ages.  Consequently,
our sample enjoys a relatively large number of such galaxies with spectroscopic
redshift measurements relative to most such studies; the spectroscopy is discussed in Paper II.

This is the first in a series of papers, which describes the target field selection, creation of K-selected sample, 
multi-band imaging, photometry, and a discussion of extremely red objects (EROs). 
Paper II discusses the spectroscopy, determination and testing of photometric redshifts, and the subsequent analysis 
of luminosity functions, dust extinctions, and star formation density.
The assumed cosmology is an $\Omega _{M}$=0.3, $\Omega _{\Lambda }$=0.7 universe 
with H$_o$=70 km s$^{-1}$ Mpc$^{-1}$.

\section{Target Fields}

Our choice of target fields is based on the presence of already known
high redshift sources, particularly where multiple detections at a similar
redshift point towards a possible over density or proto-cluster.

The correlation between associated galaxies and high redshift radio galaxies and QSOs
has been long known and established from $1<z<5$.
Many QSO metal absorption line systems are clustered on comoving scales of
$<100h^{-1} Mpc$ or less \citep{jak86,sar87,qua99}. \cite{mal96} demonstrate that targeting such metal 
absorption-line ``clusters'' yields the discovery of star-forming galaxies in emission. 
Further work \citep{tep98,man98,mey03}, has continued to discover high-redshift galaxies using this 
sort of search method.   
There is also a correlation between CIV absorbers and Lyman Break Galaxies 
\citep{ade03}, which presently make up the majority of known high-redshift galaxies.
Overdensities of
Extremely Red Objects (EROs) are found near radio galaxies \citep{cia00}
and QSOs \citep{wold03}. Recently, a cluster of H-alpha emitters was found
in association with a $z=2.16$ massive radio galaxy \citep{kurk04}; 
beyond redshift 4, Lyman alpha emitters are found to be
associated with QSOs (e.g. \citet{ven04,miley04}).

Two of our fields are selected from \cite{qua96}, who searched their database of heavy-element 
absorbers to identify nine ``superclusters'' of multiple CIV absorption systems. The QSO 0149+336 shows 
four absorbers around a redshift of $z$=2.16, while the QSO 2359+068 has two groups of six absorbers around 
the redshifts of $z$=2.83 and $z$=2.87. We chose the field near QSO 0953+549 because of multiple
known emission candidates at $z$=2.5 \citep{mal96,tep98}. Similarly, the 1714+5015 field contains 
a possible $z$=2.39 protocluster of galaxies found in WFPC2 4100\AA \ narrow band imaging \citep{pas96}. 
Finally, in an effort to sample possible over-densities throughout the whole redshift 
range of $z$=1--3, we include the target field 0741+652, which contains three emission line objects found by 
the NICMOS parallel grism survey at redshifts $z$=1.33, 1.61, \& 1.86 \citep{mcc99}. 

One field, 1107+7239, has no known high redshift sources, but has been included because of its
deep WFPC2 and NICMOS data, providing a combination of depth, photometric accuracy, and morphological 
information essentially impossible to obtain from the ground. Between the 1107+7239 and 1714+5015
fields, we have more than 100 hours of Hubble Space Telescope archival imagery;
these fields are each roughly half the exposure time of the original Hubble Deep 
Field.
The field locations, references, redshifts, and type of possible over-density are 
listed in Table 1.

\section{Optical Imaging Data}

Our optical imaging is obtained from three sources: the UCO/Lick Observatory Shane 3-m Prime Focus Camera, the Keck Observatory Low Resolution Imaging Spectrometer (LRIS), and the Hubble Space Telescope Wide Field/Planetary Camera (WFPC2). A journal of observations, listing all targets, dates, and observation times is given in Table 2.  

Optical imaging using the UCO/Lick Observatory Shane 3-m Prime Focus Camera (PFCam) was taken over 
14 nights from December 1998 to May 2002. We took deep B, V, R, and I images of 5 of 
our 6 target fields, only skipping 0953+549 because of Keck LRIS data already in hand.
For the September run in 1999, a ``Spinrad'' R filter was substituted for the more standard 
Kron-Cousins R, but that data were used only for cross-checking absolute calibration and is
not included in the final sample.
The PFCam detector is a SITe 2048$\times$2048 thinned CCD with 24-micron 
(0.296$\arcsec$) pixels, giving a field of view of 9.8$\times$9.8 arcminutes.
  
The W.M. Keck Observatory Low Resolution Imaging Spectrometer \citep[LRIS;][]{oke95} instrument has 
two detectors. The red side, or LRIS-R, consists of a 
backside-illuminated Tek 2048$\times$2048 CCD detector with an 
imaging scale of 0.215 arcsec/pixel and a field of view of 6$\times$7.3 arcmin. 
All LRIS observations 
before June 2002 and all V, R, I, and $z'$ filter observations used LRIS-R. The blue side, or 
LRIS-B, consists of two 2048$\times$4086 Marconi CCDs with extremely high sensitivity in 
the blue (Q.E. \ab 90\% at 4000\AA) with an imaging scale of 0.135 arcsec/pixel. 
While matched to the red side, the LRIS-B field of view is actually slightly larger at 
6$\times$8 arcminutes, but does contain a 13.5 arcsecond gap through its center from an 
internal bar that is blocking light. It is sensitive over the wavelength range 3100-6000 
Angstroms. All U filter observations and 2003 B filter observations were made with LRIS-B. 
The majority of the observations were made using the 560 dichroic, which splits the light 
between the two detectors at 5600\AA , although the September 2002 run used the 4600\AA\ dichroic 
and the August 2003 run used the 6800\AA\ dichroic.

Flat fields were produced from combinations of bias-subtracted twilight sky flats and images 
taken throughout the night. For I and $z'$ images, twilight flats could not be used because of 
the large spectral change in sky at those wavelengths at twilight, so flats came from images 
alone. Calibration was done by measuring multiple Landolt standard stars in a single field at 
air masses similar to the observations. For the $z'$ filter, standard star magnitudes are 
calculated using the formula of \cite{smi02}: 

\begin{equation}
z'=V - 0.841(V-R) - 1.65(R-I) + 0.51
\end{equation}

\noindent
This produces an AB magnitude which can then be converted to a Vega magnitude using the formula 
$z'$(Vega) = $z'$(AB) - 0.54.

As a consistency check, we compare the derived zero points for fields and filters in which
both LRIS and PFCam data were taken by comparing multiple (20+) moderately bright stars or 
compact objects between the two. In the ten combinations of fields and filters where this occurred, 
the difference is never more than 0.14 magnitudes and generally (7 of 10) 0.1 magnitudes or 
less. Considering the effects of different seeing and object size (there are generally few 
bright, unsaturated stars), the two sets of derived zero points appear entirely consistent. 
As a final check of the absolute calibration of our photometry, 
we compare our data to the online Sloan Digital Sky Survey Data Release 1 (SDSS DR1) catalog. 
One of our fields, 0953+549 partially overlaps with the DR1. Although their
magnitudes had to be converted from $u'g'r'i'$ to UBVRI using \cite{smi02}, we again find 
consistency to within 10\% for all filters. 

The Keck LRIS seeing ranges from 0.7-1.4 arcseconds, with the bulk around 
0.9-1.0$\arcsec$, while the Lick PFCam seeing ranges from 1.1-2.0 arcseconds, with typical 
values around 1.3-1.6$\arcsec$. Because seeing varies widely, a single aperture magnitude 
limit would be misleading. Instead the magnitude limit is a 5-$\sigma$ detection within an 
aperture 1.3 $\times$ FWHM, a size found to produce a superior signal-to-noise ratio for a typical 
compact, faint detection. Not surprisingly, the Keck LRIS magnitude limits are significantly
deeper, providing roughly a 2 magnitude advantage across all filters they share in common, 
with LRIS limits of B\ab27, V\ab27, R\ab26, and I\ab25 compared to PFCam limits of B\ab25, 
V\ab25, R\ab24, and I\ab23.5. Also, in 40 minutes on LRIS the images reach depths of 
\ab26 in U and \ab25 in $z'$ (Vega). All magnitude limit and seeing information, 
on a field by field basis, is available in Table 3.

\subsection{WFPC2 Data}

Two of our target fields contain deep pointings using the Hubble Space Telescope 
Wide Field/Planetary Camera (WFPC2). The 1714+5015 field, also referred to as the Hercules 
Deep Field, is centered on the $z=2.39$ radio galaxy 53w002 and represents 
some of the first ultra-deep imaging with the refurbished WFPC2, taken under the GO programs 
5308 and 5985 (PI: R. Windhorst) in May 1994 and June 1995. 
The deep narrow-band imaging in this field revealed 18 associated 
galaxies appearing to be at nearly the same redshift, thought to be protogalactic clumps 
that might eventually merge \citep{pas96}. The second field, 1107+7239, was observed in April 
1998 by R. Edelson and M. Malkan in parallel with a Seyfert galaxy monitoring campaign (NGC 3516; GO program 7355),
producing adjacent deep NICMOS and WFPC2 fields over 43 orbits taken within HST's continuous viewing zone. 
We present the 1107+7239 WFPC2 image in Figure 1.

The WFPC2 detector consists of four separate CCDs, the three wide field cameras (WFC1,2,3) with 
a 0.1 arcsecond pixel scale and the planetary camera (PC) with a 0.05 arcsecond pixel scale, 
which together make the distinctive chevron shape. One major drawback of the WFC chips is that 
while they have greater sensitivity than the PC chip, they undersample the image, resulting in 
a loss of resolution. To regain some of that resolution we used variable-pixel linear 
reconstruction, known more commonly as drizzling (See DITHER/DRIZZLE web page at www.stsci.edu 
and \cite{fru02}).

We drizzled all four 1107+7234 images: F300W, F450W, F606W, and the F814W filters. 
All the 1107+7234 data are composed of multiple pointings, ranging from 7-12, each with a 
small rotation in the telescope around the primary NGC 3516 STIS target, with many 
fractional pixel shifts. All drizzling was done with the IRAF/STSDAS dither package and in an 
identical manner. We aligned the images by producing a cross-correlation for the WFC chips 
with the greatest number of bright objects and then used those offsets to calculate the position 
of the other three chips. This was especially useful for the F410M PC chip image and the 
1107+7234 F300W PC, WFC1, and WFC3 images, which did not have enough signal on an individual 
image for cross-correlation. We drizzled the images from each of the detectors onto a 
single large image with a pixel scale of 0.05 arcseconds for the 1714+5015 field and 0.04 
arcseconds for 1107+7234, the difference being the higher inherent resolution of the F300W data. 
The only other difference between the reductions for the two fields was the necessity to 
correct for rotation while calculating the offsets of the 1107+7234 data.  

The F606W and F814W images from the 1714+5015 field were generated using only two pointings 
(and only one pointing in the case of the single F606W offset field) and are therefore not 
suitable for drizzling. Instead the final images were produced using standard
tools in the IRAF/STSDAS, removing cosmic rays, bad pixels (from data quality file), and 
warm pixels (hot pixels recorded by Space Telescope after every decontamination) from the 
already pipelined images before combining into a final image. In a few cases bad pixels 
overlapped, leaving a few bad spots in the final image, but generally the final image quality 
was excellent. 
We then created both drizzled and non-drizzled versions of the F450W and F410M images from their 
four separate pointings. The drizzled versions were matched to each other, while the 
non-drizzled images were matched to F606W and F814W. 

Observations of the 1714+5015 field include 16 hours of integration in F450W, 
5.7 hours in F606W, and 5.7 hours in F814W. This can be compared to 33.5 hours, 30.3 hours, 
and 34.3 hours in the same filters for the Hubble Deep Field \citep{wil96}.
We ran SExtractor V.2.0.19 \citep{ber96} on a combined image produced from a weighted average
of the three wide filter images (F450W, F606W, \& F814W), selecting objects with 5 connected pixels of 1.5-$\sigma$ 
each, except for the planetary camera (PC) chip where 9 connected pixels were required. 
Magnitudes were determined using a 0.5$\arcsec$ aperture, producing 5-$\sigma$ detections 
limits of 28.5, 28.4, and 27.1 for the F450W, F606W, and F814W filters respectively. 
Added to this are a second F606W field (a 2 hour integration) immediately south of the prime field and 
18 hours of imaging in the narrow-band filter, F410M.

Observations of the 1107+7239 field include 9 hours in F450W, 3.9 hours in F606W, 
3.9 hours in F814W, and 24.3 hours in F300W (which is almost 2/3 of the Hubble
Deep Field integration in that crucial filter). 
For this field we required 11 connected pixels 
of 1.5-$\sigma$ each, as the 1107+7234 image was drizzled, requiring more 
connected pixels to avoid spurious detections. Magnitudes were also 
determined with a 0.5 arcsecond aperture, producing 5-$\sigma$ detection limits of 
26.1, 27.9, 27.8, and 27.0 for the F300W, F606W, and F814W filters respectively.

\section{Near-Infrared Data}

The bulk of the infrared observations were done with the UCLA Gemini Twin-Arrays Infrared 
Camera \citep{mcl93} on the UCO/Lick Observatory Shane 3-m telescope. Using a dichroic 
beam-splitter, this infrared camera images in two filters simultaneously with two 
separate detectors. The short wavelength camera operates from 1-2.4 $\mu$m using a Rockwell 
NICMOS3 HgCdTe array, while the long wavelength side operates from 2-5 $\mu$m with its SBRC 
InSb array. Both detectors are 256$\times$256 pixels in size, with a scale of 0.68 
arcseconds per pixel, for a field of view nearly three arcminutes on a side. For all 
observations, either the J or H filter was used on the short wavelength side, while 
simultaneously imaging in the K' filter ($\lambda _{c}$= 2.1235, $\Delta$ $\lambda$ / 
$\lambda$ = 0.16). We carried out all observations over 13 nights from December 1998 to 
March 2002, using 9-point dithers and multiple co-adds. 

Additional observations of the 0953+549 field were made from June 3-4, 2001, at the Kitt Peak National Observatory 
(KPNO) Mayall 4-meter telescope with the NOAO Simultaneous Quad Infrared Imaging Device \citep[SQIID;][]{ell92}. 
Using a series 
of dichroics and flat mirrors, the SQIID instrument separates the incoming beam into four separate wavelength 
channels: J, H, K, and L'. Each channel illuminates a separate 512$\times$512 quadrant of a 1024$\times$1024 
ALADDIN InSb infrared array, allowing simultaneous observations in all four filters. The sky background at L', 
however, is simply too high to allow measurement of the high-redshift objects in this study.
While at the f/15 focus of the 4-m, the SQIID pixel scale is 0.39$\arcsec$ per pixel, with an 
un-vignetted field of view of 2.9$\times$3.0 arcminutes in each channel.

For infrared imaging of the 1714+5015 field we used the MDM/Ohio State Array Infrared Camera \citep[MOSAIC][]{pog98}, 
which is also referred to as TIFKAM or ONIS (Ohio State/NOAO Imaging Spectrograph), at the MDM Hiltner 
2.4-m telescope. The MOSAIC used a 512$\times$1024 
InSb array with a pixel scale of 0.30$\arcsec$ per pixel at the f/7 focus of the 2.4-m, providing a 
field of view of approximately 5.12$\times$2.56 arcminutes. The observations were made on 
seventeen nights between May 12, 1998 and July 1, 1999. We used 
three different filters: J$_{Barr}$ ($\lambda _{c}$=1.267\m, FWHM=0.271\m), H$_{Barr}$ 
($\lambda _{c}$=1.672\m, FWHM=0.274\m), and K$_{Barr}$ ($\lambda _{c}$=2.224\m, 
FWHM=0.394\m). Total integration times for the three filters were 6.18, 5.23, and 16.03 hours, respectively.
The observations are divided into two pointings, an east field and a west field, each roughly one 
arcminute apart. This gives a central region in the final image, approximately 1.8$\arcmin$ $\times$ 
5$\arcmin$ with twice the integration time of the outer regions. Combining both low and high 
signal-to-noise ratio regions gives an area of 16.2 arcmin$^{2}$ down to K=20.5 or better. 

Images were flat-fielded using a median of all data from each night, while sky frames were created
from medians of the data taken immediately before and after.
This process, similar, if more critical, to that done to the optical PFCAM data, can also lead to faint, 
negative images near bright objects in the final image. After sky subtraction and flat-fielding,
stars on each frame are examined to determine the relative signal-to-noise ratio and
seeing, which were used to produce weights for averaging, and the general photometric quality of
the data. Particularly bad frames (bad sky subtraction, a bias or dark fluctuation, tracking 
loss, cloud passing overhead, planes passing overhead, etc.) were removed. Finally, the images 
were masked for known bad pixels, shifted to align bright objects on each frame, 
and averaged together.

Calibration was done by observing faint UKIRT and \cite{per98} infrared standards throughout the 
night at air masses similar to the observations. The K' calibration is interpolated from the J, H, and K 
standard magnitudes, but it is only 0.01-0.04 magnitudes different than 
assuming K magnitudes apply. As a final check of our calibrations, we compared the photometry 
of our brightest objects against the 2 Micron All Sky Survey (2MASS) photometry maintained by 
the NASA/IPAC Infrared Science Archive (http://irsa.ipac.caltech.edu). The 2MASS
photometric catalog has 10-$\sigma$ point source detections down to at least J=15.8, H=15.1, 
and K$_{s}$=14.3 and contains sources extracted with a signal-to-noise ratio greater than 7 
in any one band or signal to noise ratio greater than 5 in any two bands. Objects
of this brightness are easily detected in our images, typically 3-4 objects per 
3$\times$3 arcminute image. Roughly 35\% of the UCLA Gemini fields (J,H, \& K' considered separately),
required minor zero point adjustments, roughly 10-15\%. 
The MOSAIC 1714+5015 images, a combination of data across eight nights, required more 
significant final corrections to the zero points:
0.32, 0.14, and 0.23 magnitudes in the J, H, and K filters respectively. These reductions show
that some non-photometric nights were erroneously considered to be photometric at the time
of the observations.

Due to unusually high background at time of observation, the SQIID data are slightly shallower than those 
taken with the UCLA Gemini camera, but with marginally better seeing: \ab 1.6$\arcsec$.
UCLA Gemini seeing is typically 1.8-2.0$\arcsec$, degrading to 2.5$\arcsec$ in a few instances. 
We combined the 0953+0549 images from the two 
instruments, weighted by signal-to-noise ratio and seeing, producing a final image with seeing and 
signal-to-noise ratio superior to the original UCLA Gemini image. The two K filters are not identical: standard K vs. 
K'. However, the estimated likely K' differences from interpolated star magnitudes are on the order 
of 1-4\%, so final averaged photometry calibration errors, where the two input images were similarly weighted, 
should be roughly half of that. This is small compared to most of the photometric errors used, and is deemed 
acceptable for the signal gain.   

Table 4 contains a summary of magnitude depths, image sizes, and seeing for all fields and partial fields.
Again, because seeing varies (1-2$\arcsec$), the magnitude limit is 
a 5-$\sigma$ detection within an aperture 1.3 $\times$ the FWHM. These are not total magnitudes,
estimation of which will be covered in more detail below.

In total, we imaged 10 partial near-infrared fields at six different target locations, covering 75.6 arcmin$^2$ 
down to a {\it total} K' magnitude of 20 and 36.3 arcmin$^2$ down to K'=20.5 or fainter.

\section{Photometry, Completeness, and Infrared Number Counts}

We extract 842 K-selected objects from the K and K' images using the SExtractor V.2.0.19 package 
\citep{ber96}, which finds and extracts objects after convolving the 
field with a Gaussian matching the measured FWHM and subtracting off the remaining background. 
 typical requirement is 9 connected pixels all above 2-$\sigma$. We change this criteria slightly from field to 
field to ensure selection of objects well below the 5-$\sigma$ aperture limits.

For photometry, SExtractor extracts both a BEST magnitude and an aperture magnitude corresponding to 
1.3$\times$FWHM. The BEST magnitude is either a Kron elliptical aperture \citep{kro80} or isophotal magnitude 
for blended/crowded objects. The object detection list for each field is then taken and cleaned of 
spurious bad pixel or noise features, objects clipped by the edges of the image, and bright 
stars. The stars are determined using the SExtractor CLASS\_STAR output parameter and comparison to the actual 
image. 
As a final cut, any detection that has an aperture magnitude greater (i.e., dimmer) than the 5-$\sigma$ 
limits quoted in Table 4 is removed. 
A distribution of all K-selected objects is shown in Figure 2.
Just from examining the turn-over in counts, we can get a rough idea that the typical data is complete 
to \ab19.5, while the deepest data are complete to \ab 20. For a more rigorous examination of 
completeness, we examine the extraction of artificial galaxies.

\subsection{Creating Artificial Galaxies To Test Completeness and Photometry Extraction}

To check completeness and the accuracy of using the SExtractor BEST magnitudes, we create artificial galaxies
using the IRAF program MKOBJECTS, available in the noao.artdata package. The program allows selection of 
ellipticity, half-light radius, profile type, and position angle. From quantitative morphological examination of 
galaxies in the Hubble Deep Field \citep{mar98}, we have detailed distributions of both half-light and 
eccentricity in the F814W filter down to 26th magnitude (AB). As a simple model, we can break both these 
distributions into two groups. Looking at half-light radius, 75\% have a small r$_{hl}$ with the distribution 
peaking around 0.25$\arcsec$ while 25\% have a larger r$_{hl}$ roughly centered at 0.5$\arcsec$. For 
eccentricity, 75\% have a small $e$ around 0.25 while 25\% have larger $e \sim$ 0.55.
Assuming no correlation between r$_{hl}$ and $e$, this simple model presents a distribution of galaxies that 
is 56\% small with low eccentricity, 19\% small with high eccentricity, 19\% large with small eccentricity, 
and 6\% large with large eccentricity. While deep {\it HST} imaging finds some differences
in morphology between galaxies imaged at 0.8\m \ and the near-infrared, such as fewer irregulars, generally
both wavelengths show a similar distribution of morphologies \citep{tep98b}. We therefore apply the F814W
models to our K' data, as the relatively minor morphology differences are
unlikely to have a significant effect on these sort of simplified distributions. 

We create four model galaxies to represent this simplified galaxy distribution: [r$_{hl},$e$]$ = [0.25,0.25], 
[0.25,0.55], [0.5,0.25], \& [0.5, 0.55]. We choose an exponential disk 
for the galaxy profile type, as evidence shows that deep galaxy counts are better fit 
by disk than by bulge models \citep{mar98}. However, as roughly 25\% of sources at these near-infrared 
depths are ellipticals \citep{cas05}, we also examine a de Vaucouleurs profile to check how 
that might change the results. Having created the models, we convolve them with a Moffat point spread function
derived from examining stars from the image fields, and scale their fluxes to a series of magnitudes from 
18-21 in 0.25 magnitude intervals. Then each model galaxy is placed into the images repeatedly at random 
locations, and extracted using SExtractor with the exact same parameters and cuts
used to extract the final K-selected object lists. This is done for every target field, as there are large 
field-to-field variations in depth and seeing.

Because of the random placement of the model galaxies, 100\% recovery (and therefore completeness), is not 
possible, even for galaxies a magnitude or two above the limits of the image. There are enough bright and/or 
large objects to hide 5-15\% of the fainter input galaxies in most images. Therefore, to find the magnitude 
limits at which completeness drops purely because of signal-to-noise ratio effects, one needs to 
look for an abrupt drop in the recovery rate, where, for example, completeness will drop 20-30\% between 
0.25 magnitude width bins. With this in mind, we record where completeness drops below 85\% for each field. 
This should roughly correspond to the magnitude depth for 100\% completeness, if one had hand-placed all the 
model galaxies in empty parts of the image. Four of the fields, 1107+7239, 1714+5015 
Center, 0149+336 Prime, and 0741+652, show 85\% completeness or better down to K=20 and 50\% completeness or 
better down to 20.25. The rest are all 85\% complete down to 19.5 or better. Completeness limits are summarized
in Table 4. Use of the de Vaucouleurs profile model galaxy produces completeness limits ranging from 0-0.5 
magnitudes brighter. 

We also examine the accuracy of the BEST magnitudes for this data. We plot the median difference in magnitude
between the input exponential disk model magnitude and its extracted BEST magnitude versus the input magnitude 
in Figure 3. This includes all fields but excludes any data from magnitude bins with less than 55\% 
completeness. The standard deviation of this difference from field to field is represented by the error bars. 
For all magnitude bins this median difference is less than 0.05 magnitudes and within one standard deviation 
of zero. We also check the de Vaucouleurs profile, which produces a 
BEST magnitude which consistently underestimates the total flux of the input magnitude by about 0.15 magnitudes. 
However, as we expect most galaxies to be better fit by exponential
disk profiles, the BEST magnitudes are an accurate measurement of total flux.

\subsection{Near-infrared Number Counts}

To produce K-band number counts, we combine all the data down to the 85\% completeness limit in each field, 
except for the four deepest fields, where we use data down to the 50\% completeness limits. Figure 4 shows 
the final K number counts for all fields, with and without completeness corrections, plotted with several 
other deep K number counts from previous studies  \citep{mcl95,min98,sar99,mai01}. 
The number counts are entirely consistent, generally agreeing well within the 
error bars, with the exception of the bright end, which is high, and a single bin at 17.25 magnitudes,
which is low. Neither are more than 2-$\sigma$ off from the rest. The data also do not support the low number
density values at the bright magnitudes from \cite{sar99}, which consistently lie below the other studies 
as well as this one.

\section{Combining Near-infrared and Optical Photometry}

All other images taken with different filters are matched to the K-band image. To accommodate different 
pixel scales, the lower resolution images are magnified to match the higher resolution images. Since most 
of the infrared data is taken at a pixel scale of 0.68$\arcsec$/pixel, this sometimes
led to large magnifications, particularly when matched to the LRIS-B 0.134$\arcsec$ or WFPC2 0.1$\arcsec$ pixel 
scale. In the case of 1107+7239, the WFPC2 data were re-sampled to match a  
coarser pixel scale (0.298$\arcsec$/pixel), as the drizzled 0.04$\arcsec$/pixel scale would have been too 
unwieldy. The data are registered to the K-band image by measuring 20-40 objects in common and fitting 
a second order polynomial to determine the needed transformation between pixel values and sky coordinates. 
 
To match photometry and produce accurate colors, all images are blurred to the same seeing FWHM by 
convolving the images with a Gaussian. While a Gaussian convolution is not a perfect representation 
of the seeing, we choose it as a better solution for measuring accurate colors than the
adoption of different sized apertures, which cannot entirely account for significant color
gradients. Deviations of the Gaussian-convolved images from those effected only by atmospheric
seeing are undetectable for small FWHM changes. This does not continue to be true, however, for
more extreme convolutions (see WFPC2 matching below).
  
Typically we blur the data to match the J image, although 
in a few cases (0953+549 \& 0741+652) the final K image has the worst seeing. One J image data from the 
southern 2359+068 field has grossly substandard seeing (3.4$\arcsec$) and is only included as no other J band 
data are available. It has been completely removed from the normal matching process, and seeing is matched to
the K image instead.

Once all data are registered, magnified, and blurred, we run SExtractor on the original non-blurred K image 
as described above and use those object detections to extract the photometry from each of the blurred images 
using an aperture magnitude corresponding to 1.3$\times$FWHM of the final, blurred image. The final magnitudes 
for each object in each filter are then defined as the K image BEST magnitude plus the 
difference between the blurred K image aperture magnitude and the blurred aperture magnitude for each filter:

\begin{equation}
 N_{final} = K_{original}(BEST) - [K_{Blurred}(Aperture) - N_{Blurred}(Aperture)]
\end{equation}

\noindent
where N is the magnitude in the filter being examined.

One exception to the above procedure is the addition of WFPC2 data to ground-based infrared and optical bands, 
as we did not wish to blur the WFPC2 images from a FWHM of roughly 0.1$\arcsec$ to 2$\arcsec$.
To measure the total WFPC2 magnitudes, we extract the BEST magnitude for all the unblurred 
F814W objects detected at greater than 5-$\sigma$ in a 0.8$\arcsec$ aperture and apply that same 0.8$\arcsec$ 
aperture to each of the remaining WFPC2 filters (F606W, F450W, and also F300W for 1107+7239), determining the 
total fluxes from the aperture differences from F814W. 
In the rare instance of multiple sources within an arcsecond of each other, we combine 
their WFPC2 fluxes into a single magnitude, as they would be indistinguishable at the resolution of the 
infrared data. This method provides accurate colors except in the extreme case of a strong intrinsic galaxy color
gradient. 

As a final step before producing colors or photometric redshifts, a correction for galactic extinction is made.
This uses the galactic extinctions of \cite{sch98} as given by the NASA/IPAC Extragalactic Database (NED). 
While the extinctions at K are essentially negligible (0.002-0.029), this is not the case with the other 
wavelength extreme, U band, for which extinction can be as high as 0.435 magnitudes. A list of galactic 
extinction E(B-V) is included in Table 1. We produced color-color diagrams of this final combined photometric
data set, which show good agreement with expected stellar tracks (see Paper II for further discussion).

\section{Extremely Red Objects}

Extremely Red Objects (EROs), first found in early deep K-band imaging \citep{els88}, are sources much
brighter in the near-infrared than in the optical. EROs are generally defined as having R-K $>$ 5, although other 
overlapping criteria (e.g., R-H$>$4, I-K$>$3.5, I$_{775}$-K$>$3.92) have also been used \citep{fir02, dad03, roc03}. 
These ERO colors are consistent with either an old, passively-evolving stellar population or 
significantly dust-extincted star formation at moderate to high redshift. 
The density and nature of these objects has been used to infer either
significant quantities of early massive galaxy formation or substantial hidden star formation \citep{cib02,miy03}. 
Complicating matters further, EROs are strongly clustered on the sky, producing a large cosmological variance from 
field to field. This variance can easily range from factors of 3-6 in surface density for even large 
near-infrared fields of view \citep[25 arcmin$^2$;][]{dad00}.  

We examine this effect by identifying EROs within 5 of our 6 near-infrared fields.  We exclude
0149+336 from this analysis because of its relatively shallow optical (Lick only) imagery.
The remaining sample contains 481 objects down to a K magnitude of 19.7, covering 61 square arcminutes. We plot R-K versus K in Figure 6. For most fields the filter plotted is actually K', which will have a negligible impact on the R-K color, unless a break were to fall within the K filter (z\ab 4 for the 4000\AA\ break). All fields use the same R filter from the Keck LRIS.

We find 95 EROs (R-K $>$ 5; K $<$ 19.7) in our sample for a density of 1.55$\pm$ 0.16 arcmin$^{-2}$. 
This compares well with the density of $1.60\pm 0.09$ found by  \citet{george05}, adopting
$I-K>4$ and $K<20$ in the Phoenix Deep field.
The density changes significantly across the five fields, ranging from 0.67-3.3 arcmin$^{-2}$, with the largest deviation being the 1714+5015 field which contains slightly more than half (52) of all candidates in only 25\% of the observed area. Excluding 1714+5015 produces a density of 0.95$\pm$ 0.14 arcmin$^{-2}$. A fluctuation of 5 between fields is within the range of predicted density fluctuations. The 1714+5015 field K-band number counts show no large disparities, within the errors, from the K-band number counts of all the combined fields. The 1714+5015 counts do seem to be slightly higher from K=18.25-19.5, but this is of marginal significance. There is no significant spatial clustering visible within
the 3$\times$5 arcminute near-infrared 1714+5015 field image, nor within the other smaller fields with lower ERO density.  
If we place the magnitude cut-off at K=19.2 to allow easier comparison with previous studies, we measure a density of 1.10$\pm$ 0.13 arcmin$^{-2}$. This agrees reasonably well with most previous studies \citep{dad00,cib02,fir02} that find densities of 0.5-1 EROs arcmin$^{-2}$, where our result is slightly high because of the single extreme 1714+5015 field.
 
A high density of red objects has been previously noted within the 1714+5015 field, with
\cite{im04} reporting a population of Hyper-EROs (HEROs), objects with J-K $>$ 2.7 believed
to be at redshifts of $z>2$ \citep{im02}.
However, these objects are in general not the EROs we report here, as the HERO population are mostly 
much fainter (K$>$20.5; Im, M. 2005, private communication). Our study is not sensitive to anything 
but the brightest HEROs, as much deeper J-band data are required to confirm such extreme near-infrared colors.  
We further investigate the possibility that the 1714+5015 EROs might be at very high redshift by performing a 
B-z versus z-K color-color (BzK) analysis, which is capable of selecting both star-forming and passive galaxies 
at $z>1.4$ \citep{dad04}. We find only 30\% of the 1714+5015 ERO sample is above $z=1.4$, evenly distributed between 
star-forming and old galaxies, which indicates the majority of the EROs are not associated with the known
$z$=2.4 over-density \citep{pas96}.
  
We examined the morphologies of those EROs that lay within the 1714+5015 WFPC2 F814W images (12 total). 
Because they are very faint at optical wavelengths we did not attempt to extract any morphological parameters, as 
meaningful results require signal-to-noise ratios of 50-100 \citep[ex.,][]{con03}. The majority (10 of 12)
are extremely compact, which we subjectively split into 6 disk-like, 3 spheroid-like, 
and 3 so compact as to be essentially point-like (although they are resolved). Of the remaining two EROs, 
one is clearly an elliptical, while the other is an extended disk with arm-like features. 

Spectroscopic redshifts were obtained for 18 of the 95 EROs (See Paper II for spectroscopy details), which 
ranged from z=0.53-2.55 with most lying in the z=0.7-1.2 range. None are obvious AGNs or stars. All but one 
were identified from emission lines, making them almost all dusty starbursts. This is almost certainly a strong 
selection effect, as emission spectra are more easily identifiable than absorption spectra, particularly for 
the reddest objects in our sample.
We failed to obtain spectroscopic redshifts for an additional 24 K$<$19.7 EROs, indicating as much as 60\% 
of the ERO sample could be passively evolving. 
Previous studies have indicate a ratio of passively evolving to dusty star-forming EROs close 
to 1 \citep[i.e.,][]{cib02,miy03}.   
No significant redshift clustering is seen in the ten redshifts measured for the 1714+5015 field to explain the ERO 
over-density, although there are three different pairs of galaxies at the same redshift (z=0.90,z=1.03,z=1.22), 
possibly indicating a chance alignment of smaller over-densities.

\section{Summary}
 
This paper presents data from the Hubble Space Telescope, three 
ground observatories, five different telescopes, seven different instruments, eight
different fields, and fourteen different wavelength filters, covering the 
ultraviolet through near-infrared. The combined set of optical and near-infrared photometry,
roughly 75 arcmin$^2$ down to K=20-20.5, consists of six different fields 
throughout the sky, greatly reducing the vulnerability to cosmic variance.

From these target fields we extract 842 K-selected sources, producing number counts consistent with
previous near-infrared samples. After producing optical through near-infrared colors, we measure
an ERO density of 1.55$\pm$ 0.16 arcmin$^{-2}$, higher than most previous studies, but within the known variance 
of near-infrared fields of the size examined. Roughly half of the detected EROs lie within the 1714+5015 field,
a field already known to contain an over-density of red galaxies. Spectroscopic confirmations indicate
the explanation for the high ERO density in 1714+5015 may not be a single cluster, but a chance alignment of 
red galaxies at multiple redshifts. 

Most major studies of the high-redshift universe have been done in 
the rest-wavelength ultraviolet, potentially missing a significant redder
high-redshift population. Our K-band selected sample allows examination of these redder galaxies out to
the Bright Ages ($z$=1-2.5).
In Paper II we will discuss the follow-up spectroscopy and detection of a significant number
of galaxies at high redshift, allowing the determination of reliable photometric redshifts for study 
of the evolution of luminosity functions, dust extinction, and star formation density.

We acknowledge assistance from the team that obtained the parallel HST observations of the 1107+7239 field,
including the Principal Investigator (R. Edelson), and A. Koratkar for the scheduling and R. Fruchter
for a preliminary reduction of some of the data. We would also like to thank Myungshin Im for his valuable input.
We am grateful for the work of Ian McLean and the entire UCLA Gemini Twin-Arrays Infrared Camera team, 
without which this project would not have been possible. We wish to also acknowledge the use of TIFKAM, which 
was funded by the Ohio State University, the MDM consortium, MIT, and NSF grant AST-9605012. NOAO and USNO
paid for the development of the ALADDIN arrays and contributed the array currently in use in TIFKAM. 
The assistance of the staffs of the Lick, Kitt Peak, and Keck observatories was invaluable. Finally, 
We would like to acknowledge financial support from a grant for HST program AR-9543.

\begin{deluxetable}{lllcccc}
\setlength{\tabcolsep}{0.05in} 
\tablehead{
\colhead{Target Field} &
\colhead{RA} &
\colhead{Dec} &
\colhead{Gal. Ext.} &
\colhead{\it z} & 
\colhead{Type} &
\colhead{Ref.} \\ 
&
\colhead{(J2000)} &
\colhead{(J2000)} &
\colhead{E(B-V)} & & &  
}
\tablecaption{Target Fields}
\startdata
0149+336  & 01$^h$52$^m$34.6$^s$  & +33$\deg$50$\arcmin$33$\arcsec$ & 0.041 & 2.16  & absorption & A  \\
0741+652  & 07$^h$41$^m$45.4$^s$  & +65$\deg$15$\arcmin$36$\arcsec$ & 0.033 & 1.33,1.61,1.86 & emission   & B \\
0953+549  & 09$^h$57$^m$14.7$^s$ & +54$\deg$40$\arcmin$18$\arcsec$ & 0.007 & 2.50  & emission   & C  \\
1107+7239 & 11$^h$06$^m$50.0$^s$  & +72$\deg$39$\arcmin$00$\arcsec$ & 0.043 & n/a & deep field & D \\
1714+5015 & 17$^h$14$^m$14.7$^s$  & +50$\deg$15$\arcmin$30$\arcsec$ & 0.022 & 2.39  & emission & E \\ 
2359+068  & 00$^h$01$^m$40.6$^s$ & +07$\deg$09$\arcmin$54$\arcsec$ & 0.058 & 2.83,2.87  & absorption & A \\
\enddata
\tablerefs{A) \cite{qua96} B)  \cite{mcc99} C) \cite{tep98} D) Deep NICMOS and WFPC2 fields taken in parallel with a STIS observation program of NGC 3516 by Rick Edelson. E) \cite{pas96}}
\end{deluxetable}

\begin{deluxetable}{llccl}
\tablecaption{\bf Journal of Observations} 
\tablehead{
\colhead{Target Field} &
\colhead{Filter} &
\colhead{Int. Time} &
\colhead{Telescope} &
\colhead{Observation Dates} \\
& & \colhead{(Seconds)} & &
}
\startdata 
0149+336 & B & 9000 & Lick 3m & Oct 2 2000 \\
& V & 9000 & Lick 3m & Oct 1 2000 \\
& R & 6600 & Lick 3m & Oct 2,3 2000 \\
& I & 7200 & Lick 3m & Oct 1,3 2000 \\
& J & 18360 & Lick 3m & Nov 16,17 2000; Jan 31 2002 \\
& K' & 19152 & Lick 3m & Nov 16,17 2000; Jan 31 2002 \\
0741+652 & U & 2400 & Keck & Jan 5 2003 \\
& V & 1800 & Lick 3m & Dec 18 1998 \\
& R & 1300 & Keck & Jan 5 2003 \\
& I & 5400 & Lick 3m & Dec 18 1998; Jan 19 2002 \\
& {\it z'} & 1060 & Keck & Jan 5 2003 \\
& J & 13320 & Lick 3m & Jan 1 1999; Jan 31 2002 \\
& H & 6960 & Lick 3m & Jan 30 2002 \\
& K' & 20306 & Lick 3m & Jan 1 1999; Jan 30,31 2002 \\
0953+549 & U & 2600 & Keck & June 11 2002 \\
& B & 1800 & Keck & Jan 5 1997 \\
& V & 2000 & Keck & Dec 24 1995 \\
& R & 1200 & Keck & June 11 2002 \\
& I & 3200 & Keck & Dec 23 1995; June 11 2002 \\
& J & 35280 & Lick 3m & Mar 19,21+Nov 16,17 2000 \\
& J & 5640 & Kpno 4m & June 3,4 2001 \\
& H & 5340 & Kpno 4m & June 3,4 2001 \\
& K'& 35060 & Lick 3m & Mar 19,21+Nov 16,17 2000 \\
& K & 5460 & Kpno 4m & June 3,4 2001 \\
1107+7239 & B & 1680 & Keck & Mar 25 2003 \\
& V & 9000 & Lick 3m & Dec 17 1998 \\
& R & 1500 & Keck & Mar 25 2003 \\
& I & 7200 & Lick 3m & Dec 18 1998 \\
& J & 12800 & Lick 3m & Jan 1, Mar 1 1999 \\
& H & 11988 & Lick 3m & Jan 1, Mar 1 1999 \\
& K'& 24336 & Lick 3m & Jan 1, Mar 1 1999 \\ 
1714+5015 & U & 6650 & Keck & Sept 13,14 2002; Feb 25 2003 \\
& B & 1200 & Keck & Feb 25 2003 \\
& V & 2200 & Keck & Sept 13,14 2002 \\
& R & 1200 & Keck & Sept 14 2002 \\
& I & 1400 & Keck & Feb 25 2003 \\
& {\it z'} & 2250 & Keck & Feb 25 2003 \\
& J & 22260 & MDM 2.4m & May 16,17,19,23,26 1998; June 19+July 1 1999 \\
& H & 18840 & MDM 2.4m & May 23,26+Oct 8,9,11 1998; July 1 1999 \\  
& K' & 57720 & MDM 2.4m & May 12,14,15,23,24,26 1998; June 18,20 1999 \\
2358+068 & U & 2650 & Keck & Aug 28 2003 \\
& B & 9000 & Lick 3m & Sept 3 1999 \\
& V & 9000 & Lick 3m & Sept 7 1999 \\
& R & 1000 & Keck & Aug. 28 2003 \\
& I & 1000 & Keck & Aug. 28 2003 \\
& {\it z'} & 1450 & Keck & Aug 29 2003 \\ 
& J & 13320 & Lick 3m & Sept 17 1999; Nov 17 2000 \\
& K & 13176 & Lick 3m & Sept 17 1999; Nov 17 2000 \\
\enddata
\end{deluxetable}

\begin{deluxetable}{lcccccccccccc}
\rotate
\setlength{\tabcolsep}{0.05in} 
\tablecaption{Ground-based Optical Field Depths and Seeing}
\tablehead{
\colhead{Optical} &
\colhead{U 5$\sigma$} & 
\colhead{B 5$\sigma$} &
\colhead{V 5$\sigma$} &
\colhead{R 5$\sigma$} &
\colhead{I 5$\sigma$} &
\colhead{{\it z} 5$\sigma$} &
\colhead{U ($\arcsec$)} &
\colhead{B ($\arcsec$)} &
\colhead{V ($\arcsec$)} &
\colhead{R ($\arcsec$)} &
\colhead{I ($\arcsec$)} &
\colhead{{\it z} ($\arcsec$)} \\ 
\colhead{Field} & 
\colhead{Lim.} & 
\colhead{Lim.} &
\colhead{Lim.} &
\colhead{Lim.} & 
\colhead{Lim.} &
\colhead{Lim.} &
\colhead{Seeing}&
\colhead{Seeing}&
\colhead{Seeing}&
\colhead{Seeing}&
\colhead{Seeing}&
\colhead{Seeing}

}
\startdata
0149+336  & n/a & 25.1 & 25.1 & 24.1 & 23.3 & n/a & n/a & 1.5 & 1.3 & 1.3 & 1.2 & n/a \\
0741+652  & 26.1\tablenotemark{a} & n/a & 23.5 & 26.0\tablenotemark{a} & 23.6 & 23.6\tablenotemark{a} & 1.4\tablenotemark{a} & n/a & 1.3 & 1.3\tablenotemark{a} & 1.3 & 1.3\tablenotemark{a} \\
0953+549  & 25.8\tablenotemark{a} & 27.1\tablenotemark{a} &  27.3\tablenotemark{a} & 26.2\tablenotemark{a} & 25.0\tablenotemark{a} & n/a & 1.1\tablenotemark{a} & 1.1\tablenotemark{a} & 0.8\tablenotemark{a} & 1.0\tablenotemark{a} & 0.9\tablenotemark{a} & n/a  \\
1107+7239 & n/a & 26.9\tablenotemark{a} & 24.4 & 25.8\tablenotemark{a} & 23.7 & n/a & n/a & 1.1\tablenotemark{a} & 1.7 & 1.2\tablenotemark{a} & 1.1 & n/a \\
1714+5015 & 26.2\tablenotemark{a} & 26.9\tablenotemark{a} & 26.4\tablenotemark{a} & 25.8\tablenotemark{a} & 25.5\tablenotemark{a} & 25.0\tablenotemark{a} & 0.9\tablenotemark{a} & 0.9\tablenotemark{a} & 0.9\tablenotemark{a} & 0.9\tablenotemark{a} & 0.8\tablenotemark{a} & 0.7\tablenotemark{a} \\
2359+068 & 25.8 & 24.8 & 24.6 & 26.0 & 25.2 & 24.5 & 0.9 & 1.6 & 1.6 & 0.9 & 0.8 & 0.7 
\enddata
\tablenotetext{a}{Keck LRIS data}
\end{deluxetable}

\begin{deluxetable}{lccccccccc}
\tabletypesize{\footnotesize }
\setlength{\tabcolsep}{0.02in} 
\tablecaption{Infrared Field Depths, Sizes, and Seeing}
\tablehead{
\colhead{Infrared} &
\colhead{J 5-$\sigma$} & 
\colhead{H 5-$\sigma$} &
\colhead{K' 5-$\sigma$\tablenotemark{a}} &
\colhead{Size} &
\colhead{J ($\arcsec$)} &
\colhead{H ($\arcsec$)} &
\colhead{K'($\arcsec$)\tablenotemark{a}} &
\colhead{K 85\%} &
\colhead{Observa-} \\
\colhead{Field} & 
\colhead{Limit} & 
\colhead{Limit}&
\colhead{Limit} &
\colhead{(arcmin$^2$)} &
\colhead{Seeing}&
\colhead{Seeing}&
\colhead{Seeing}&
\colhead{Complete} &

\colhead{tory} 
}
\startdata
0149+336 Prime  & 22.04 & n/a & 20.88 & 7.6\tablenotemark{b} & 2.1 & n/a & 1.9 & 20.0 & Lick 3m \\
0149+336 West   & 21.64 & n/a & 20.30 & 6.4 & 1.9 & n/a & 1.9 & 19.5 & Lick 3m \\
0741+652        & 22.69 & 21.63 & 20.97 & 7.1 & 1.6 & 1.7 & 1.8 & 20.0 & Lick 3m \\
0953+549 North  & 22.31 & 20.94 & 20.35 & 8.1 & 1.9 & 1.4 & 2.4\tablenotemark{c} & 19.5 & Lick 3m\tablenotemark{c} \\
0953+549 South  & 21.70 & 20.47 & 20.34 & 7.5 & 1.9 & 1.3 & 1.8\tablenotemark{c} & 19.5 & Lick 3m\tablenotemark{c} \\
1107+7239       & 21.67 & 20.73 & 20.74 & 7.7 & 2.4 & 1.9 & 2.0 & 20.0 & Lick 3m \\
1714+5015 Cent. & 22.16 & 20.94 & 20.89 & 8.8 & 1.2 & 1.2 & 1.1 & 20.0 & MDM 2.4m \\
1714+5015 Sides & 21.85 & 20.52 & 20.52 & 7.4 & 1.2 & 1.2 & 1.1 & 19.5 & MDM 2.4m \\
2359+068 North  & 21.67 & n/a & 20.27 & 7.4 & 2.2 & n/a & 2.0 & 19.75 & Lick 3m \\
2359+068 South  & 21.16 & n/a & 20.46 & 7.6 & 3.4 & n/a & 2.0 & 19.75 & Lick 3m 
\enddata
\tablenotetext{a}{Magnitudes are standard K, not K' for 1714+5015.}
\tablenotetext{b}{Of the total 7.6 arcmin$^2$, 2.3 arcmin$^2$ are actually lower signal-to-noise ratio data, roughly 
identical in depth to the 0149+366 West data. Only 5.3 arcmin$^2$ are at the magnitude limits reported for 
0149+366 Prime.}
\tablenotetext{c}{Data also taken at KPNO 4m. See near-IR section for details.}
\end{deluxetable}

\begin{figure}
  \includegraphics[width=1.0\textwidth]{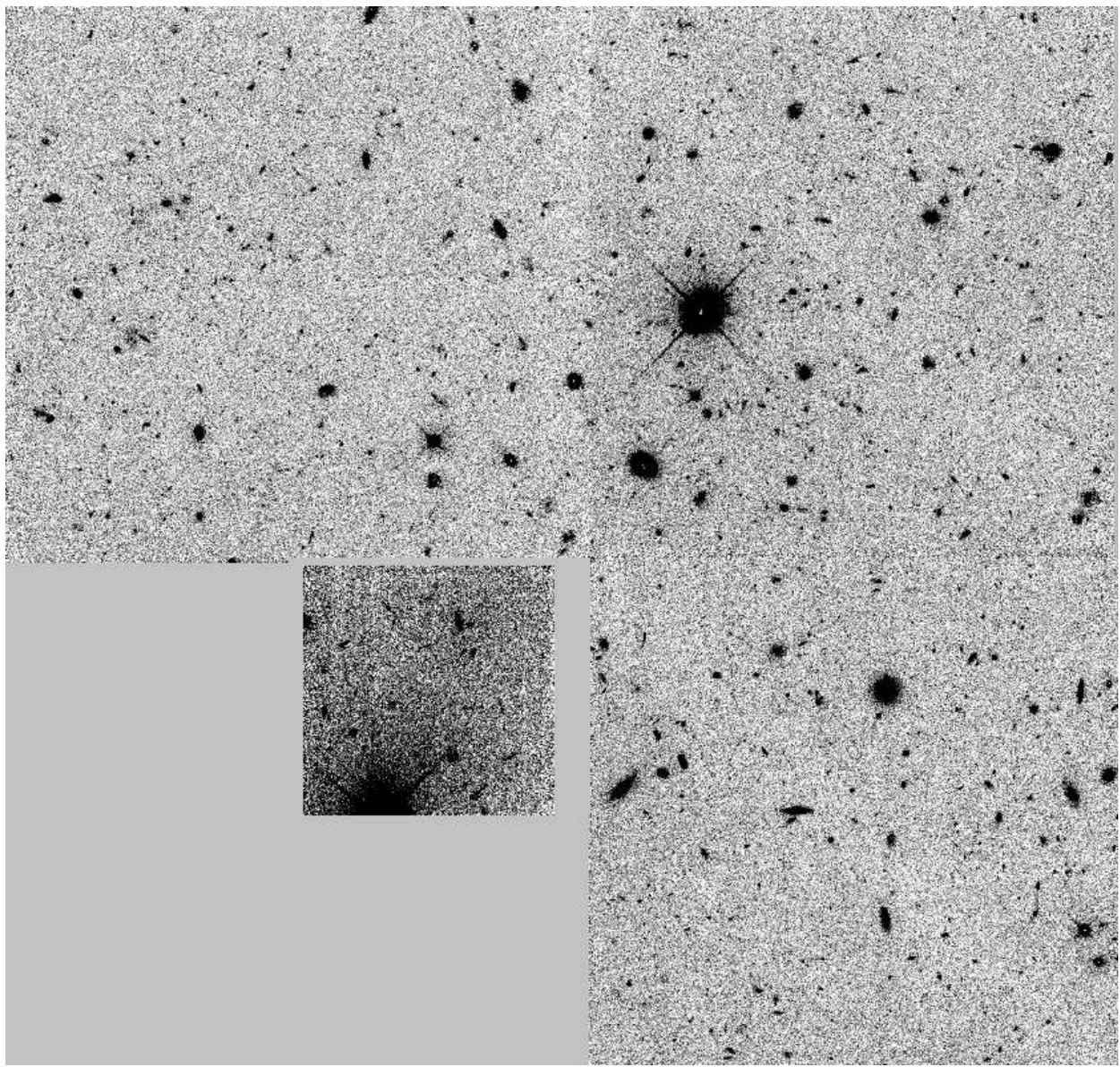}
  \caption{Greyscale F606W image of the  1107+7239 field.}
\end{figure}

\begin{figure}
  \includegraphics[angle=90,width=1.0\textwidth]{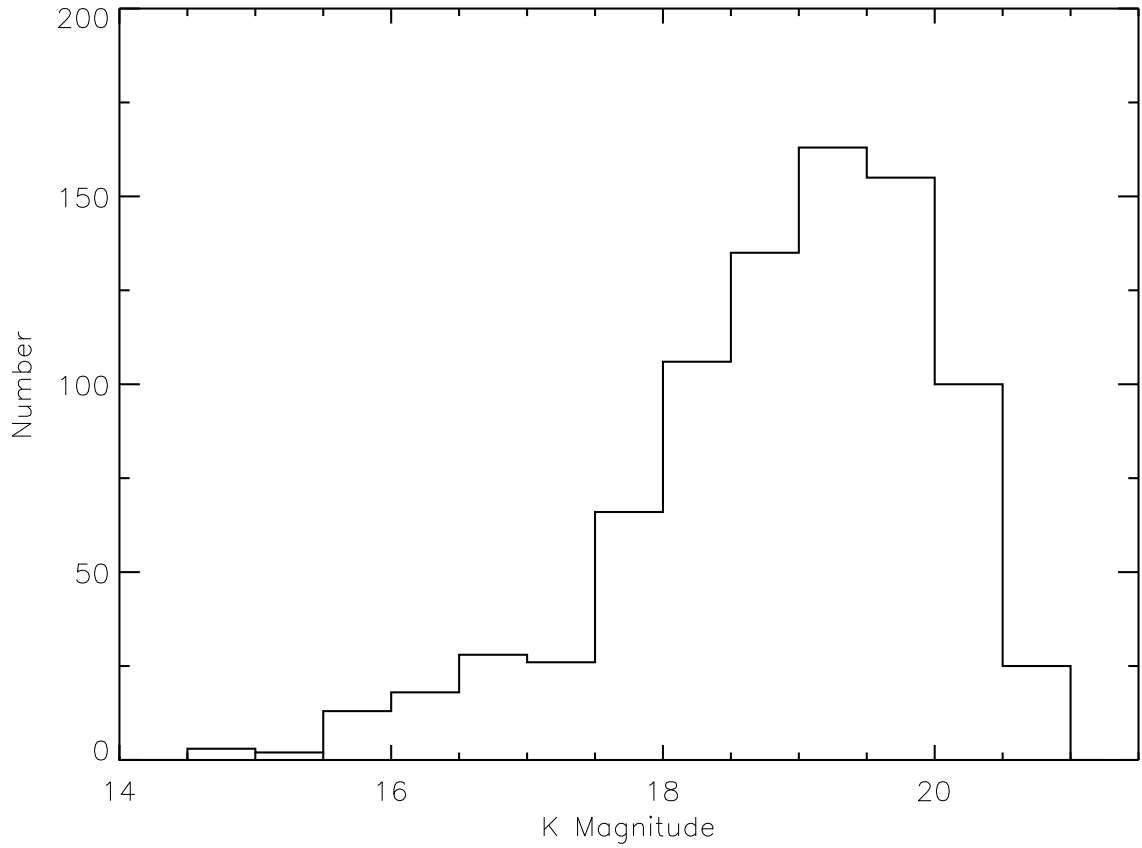}
  \caption{A histogram of all K-selected objects (including objects
later classified as stars) from the 6 fields comprising the 75.6 sq.
arcmin of this survey.}
\end{figure}

\begin{figure}
  \includegraphics[width=1.0\textwidth]{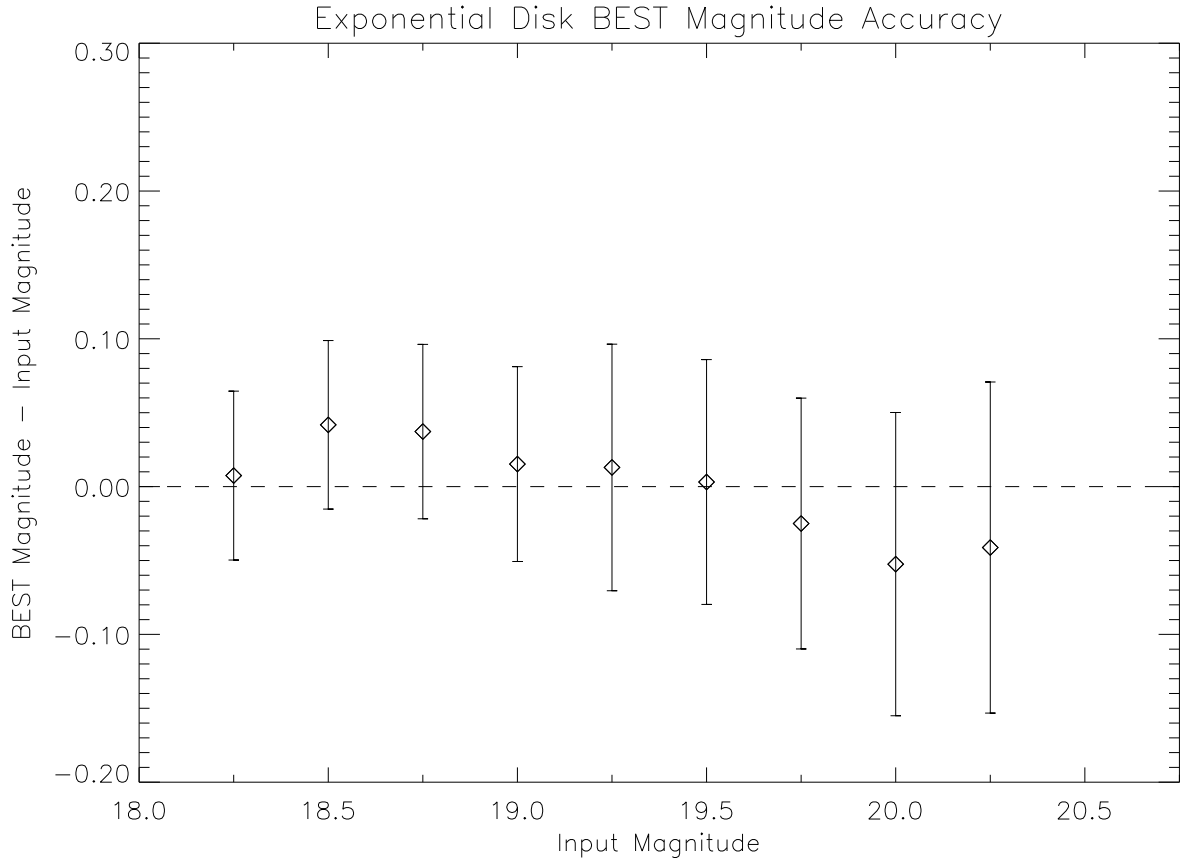}
  \caption{A distribution of the difference between the extracted BEST magnitude used by SExtractor and the 
 input magnitude of the artificial galaxies created using the exponential disk model, given as a function of the model 
 input magnitude. This plot shows that the BEST magnitude is accurate down to almost 50\% completeness, where the
 average field tested reaches K=20.25.}
\end{figure}  

\begin{figure}
  \includegraphics[angle=90,width=1.0\textwidth]{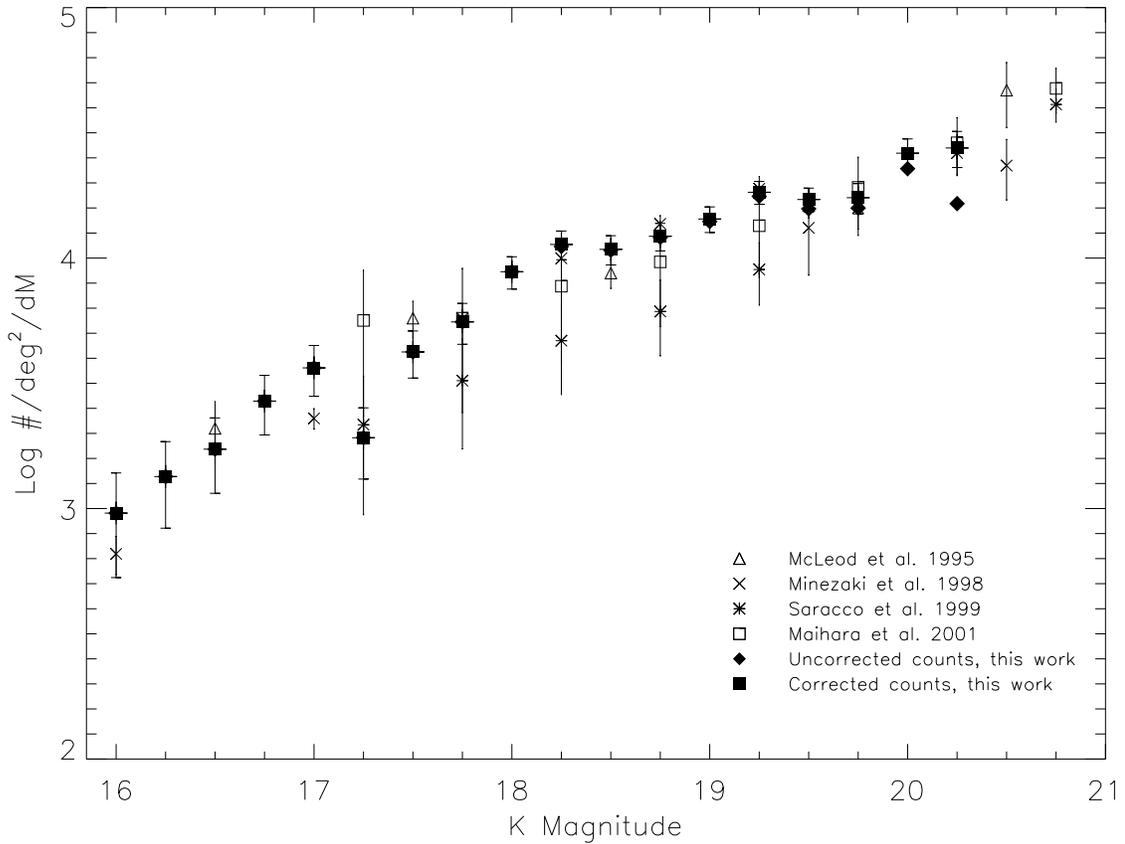}
  \caption{K-band number counts for our infrared fields. Solid diamonds are our counts, not corrected for completeness; 
   the solid rectangles are the corrected counts. Notice that completeness
   corrections are significant only for K$>$20 mag. For comparison, K number counts from 
   previous studies are also displayed: the open triangles are from \cite{mcl95}, the X's are from \cite{min98}, 
   the asterisks are from \cite{sar99}, and the open squares come from \cite{mai01}. }
\end{figure}

\begin{figure}
   \includegraphics[angle=90,width=1.0\textwidth]{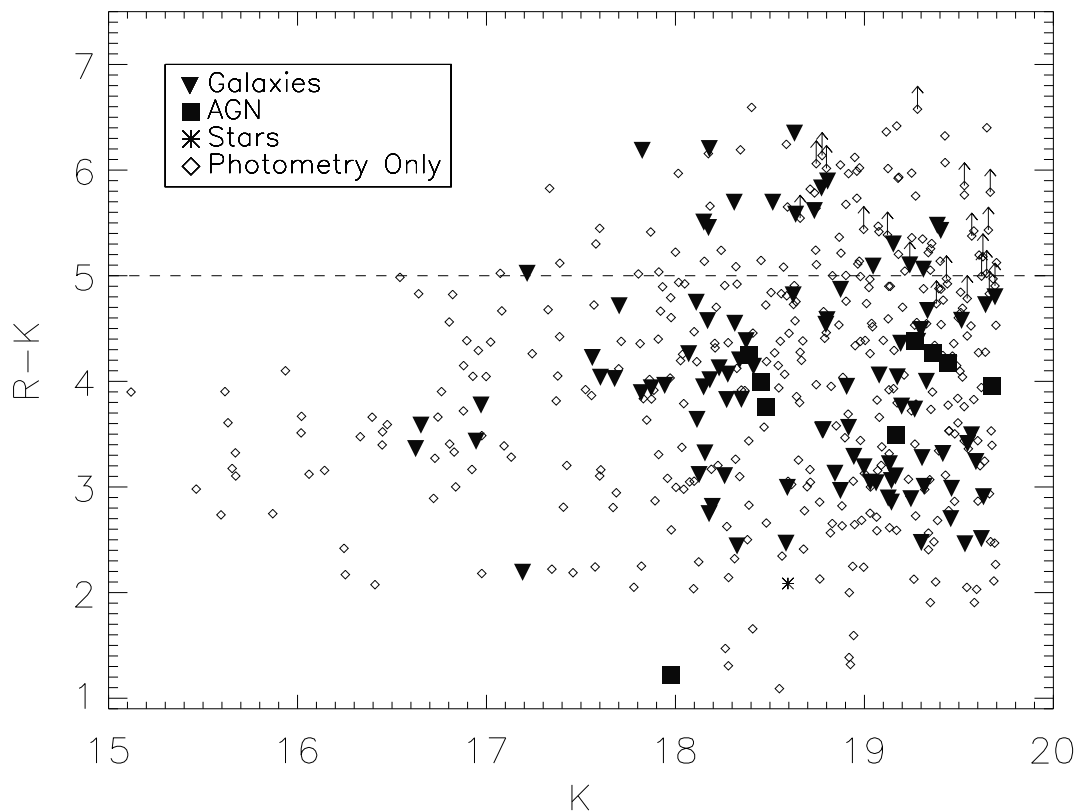}
  \caption{R-K color plotted against K magnitude for all objects with K$>$19.7, excluding field 0149+336 due to the relative shallowness of its optical data. The R-K$>$5 ERO demarcation is drawn as a dashed horizontal line. Lower limits in R-K are indicated with upward pointing arrows. Sources with spectroscopic redshifts are plotted as solid upside down triangles, AGN are plotted as solid squares, and one star (remaining after an initial color cut to remove stars) is plotted as an asterisk.}
\end{figure}

\end{document}